# An Exact Evaluation Of The Casimir Energy In Two Planar Models

S.G.Kamath [*]

Department of Mathematics, Indian Institute of Technology Madras, Chennai 600 036, India

[*]e-mail: kamath@iitm.ac.in

**Abstract**: The method of images is used to calculate the Casimir energy in Euclidean space with Dirichlet boundary conditions for two planar models, namely: i. the non-relativistic Landau problem for a charged particle of mass m for which - irrespective of the sign of the charge - the energy is negative, and ii. the model of a real, massive, noninteracting relativistic scalar field theory in 2 + 1 dimensions, for which the Casimir energy density is non-negative and is expressed in terms of the Lerch transcendent $\Phi\left(\lambda, n, \frac{1}{2}\right)$ and the polylogarithm $Li_n(\lambda)$ with $0 < \lambda < 1$ and n = 2, 3.

## Introduction

This paper follows the work of Brown and Maclay[1] to calculate the Casimir[2] energy density from the stress – energy tensor $T^{\mu\nu}(x)$ in the parallel plate configuration in 3 + 1 dimensions. Of interest here is their use of the method of images[3] as a technique to obtain for the Lagrangian density $L = -\frac{1}{4} F_{\alpha\beta} F^{\alpha\beta}$ at zero temperature,

$$\langle T^{\mu\nu}(x) \rangle_{(0)} = \frac{\pi^2}{180} \frac{\hbar c}{a^4} \left( \frac{1}{4} g^{\mu\nu} - \hat{z}^\mu \hat{z}^\nu \right) \qquad (1)$$

with the unit four vector $\hat{z}^\mu = (0,0,0,1)$, $g^{\mu\nu} = diag(-1,1,1,1)$, the two perfectly conducting, parallel, infinite plates being placed at z = 0 and z = a respectively. Note that $\langle T^{\mu\nu}(x) \rangle_{(0)}$ is homogeneous of degree $n = -4$ in the parameter 'a' ; an answer which involves more than *one* parameter besides the separation between the two plates will clearly provide variety and this is easily got by working with the Lagrangian density for a free massive real scalar field φ (x) in Euclidean space namely,

$$L = \frac{1}{2} \partial_\mu \varphi \partial_\mu \phi + \frac{1}{2} m^2 \phi^2 \qquad (2)$$

the Casimir energy density for which will be shown to be non-negative(see Eq.(10) below). As a prelude to the above exercise we first take up the calculation of the Casimir energy for the Landau problem, with the Lagrangian given in Euclidean space by

$$L_E = \frac{1}{2}m\left(\frac{d\vec{r}}{ds}\right)^2 - \frac{iq}{c}\left(\frac{d\vec{r}}{ds}\right)\cdot\vec{A}(\vec{r}), \quad 2\vec{A} = \vec{B}\times\vec{r} \tag{3}$$

In giving Eq.(3) this priority we are motivated by : i. this is a facet of the Landau problem that to our knowledge has not been reported in the literature[4] and ,ii. the results obtained below are not only pleasing, but *timely* in that Casimir forces are now believed to play a significant role[5] in micro- and nanometer size structures and hence need to be factored in the design and modeling of Micro-Electromagnetic-Systems( MEMS) and Nano-Electromagnetic-Systems(NEMS). Eq.(3) yields the Hamiltonian $H_E = \frac{1}{2}m\sum_{i=1}^{2}\dot{x}_i\dot{x}_i, \dot{x}_i \equiv \frac{dx_i}{ds}$ and the Casimir energy can now be worked out from its definition[2] as a difference in zero-point energy, namely, $E_{vac}[\partial\Gamma] = \langle 0|H_E|0\rangle_{\partial\Gamma} - \langle 0|H_E|0\rangle_0$. Following Plunien et al.[3], each matrix element in $E_{vac}[\partial\Gamma]$ can now be expressed in terms of the propagator $D_{ij}(s-s') = \langle 0|T(x_i(s)x_j(s'))|0\rangle$, with that of the second term for example, being given by

$$\langle 0|H_E|0\rangle_0 = \frac{1}{2}m\sum_{i,j=1}^{2}\delta_{ij}\partial_s\partial_{s'}D_{ij}(s-s')\bigg|_{s=s'} \tag{4}$$

$D_{ij}(s-s')$ itself is determined following standard methods [6] and it works to ( with $\varepsilon_{12} = 1$ )

$$D_{ij}(s-s') = \frac{c}{2qB}e^{-\frac{qB}{mc}|s-s'|}\left(\delta_{ij} - 2i\varepsilon_{ij}e^{\frac{qB}{2mc}|s-s'|}\sinh\frac{qB}{2mc}(s-s')\right) \tag{5}$$

The counterpart of (5) when Dirichlet boundary conditions are imposed at s = 0 and s = 2a as in Fig.1 , is easily derived from Plunien et al.[3] and Morse and Feshbach [3] and it is given by

$$\tilde{D}_{ij}(s-s') = \sum_{k=-\infty}^{\infty}D_{ij}(s-s'-4ak) - \sum_{k=-\infty}^{\infty}D_{ij}(s+s'-4ak) \tag{6}$$

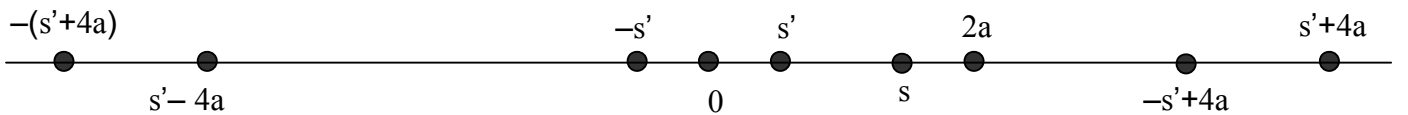

**FIGURE 1**: The image-source construction for the constrained Green's function with Dirichlet boundary conditions at s = 0 and s = 2a (a > 0).

With the label $\tau_{ij,ss'} \equiv \frac{1}{2}m\sum_{i,j=1}^{2}\delta_{ij}\partial_s\partial_{s'}$, we now write $E_{vac}[\partial\Gamma]$ as

$$E_{vac}[\partial\Gamma] = \left(\tau_{ij,ss'}\sum_{k=-\infty}^{\infty}{}'D_{ji}(s-s'-4ak)\right)\bigg|_{s=s'} - \left(\tau_{ij,ss'}\sum_{k=-\infty}^{\infty}D_{ji}(s+s'-4ak)\right)\bigg|_{s=-s'} \quad (6a)$$

The first and second terms yield $-\dfrac{qB}{mc}\dfrac{1}{e^{4a\frac{qB}{mc}}-1}$ and $-\dfrac{qB}{2mc}\dfrac{e^{4a\frac{qB}{mc}}+1}{e^{4a\frac{qB}{mc}}-1}$; thus,

$$E_{vac}[\partial\Gamma] = -\frac{qB}{2mc}\frac{e^{4a\frac{qB}{mc}}+3}{e^{4a\frac{qB}{mc}}-1} \quad (7)$$

The Casimir energy is therefore negative for either sign of the charge q. To conclude, we should reiterate that the above calculation is just another aspect of the Landau problem that has otherwise received considerable attention in the literature[4]; to recall just one of them here, the quantum energy levels are known to be infinitely degenerate and as shown by Fubini [4] is due to the invariance of the quantum Hamiltonian under "magnetic translations" though translational symmetry in the usual sense is broken.

## The Scalar Field Theory

For the Lagrangian density given by (2) the propagator is

$$\langle 0|T^*(\phi(x)\phi(y))|0\rangle = D(x-y) = \int_p \frac{e^{-ip\cdot(x-y)}}{p^2+m^2} = \frac{1}{4\pi\rho}e^{-m\rho} \quad (8)$$

with $\rho^2 = (x-y)^2$ and $\int_p$ being a label for $\int\frac{d^3p}{(2\pi)^3}$. Using Fig.2, the counterpart of Eq.(6a) but with a pair of plates at y = - a and y = + a on each of which a Dirichlet boundary condition is enforced, is now given by

$$\Theta^{\mu\nu}_{vac}[\partial\Gamma] = \frac{1}{4\pi}\left(\partial^{\mu}_{(x)}\partial^{\nu}_{(y)} - \frac{1}{2}g^{\mu\nu}\left(\partial^{\alpha}_{(x)}\partial^{\alpha}_{(y)} + m^2\right)\right)\left\{\sum_{k=-\infty}^{\infty}\frac{1}{\omega_k}e^{-m\omega_k} - \frac{1}{\omega_0}e^{-m\omega_0} - \sum_{j=-\infty}^{\infty}\frac{1}{\eta_j}e^{-m\eta_j}\right\}\Bigg|_{x=y} \quad (9)$$

with $\omega_k^2 = (x-y_k)^2$, $\eta_j^2 = (x-y_j)^2$, $y_k = (y_0, y_1, y_2 + 4ak)$ & $y_j = (y_0, y_1, -y_2 + 2a(2j-1))$.

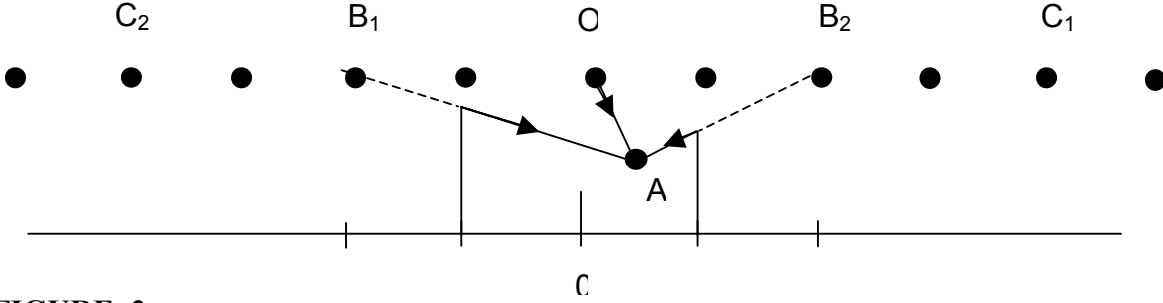

**FIGURE .2** : Image-source construction for the constrained Green's function for the parallel plate configuration at y = - a and y = + a in Euclidean space. The y - coordinate of O is zero, of $B_i$ are 2 a i, $(i = \pm 1, \pm 3, \pm 5,....)$ and that of the $C_j$ are 4 a j, $(j = \pm 1, \pm 2, \pm 3, \pm 4,.....)$ .

The first term in (9) becomes $\frac{2}{4\pi}\sum_{k=1}^{\infty}\frac{e^{-4amk}}{(4ak)^3}\left[(1+4amk)(g^{\mu\nu} - 3\hat{y}^{\mu}\hat{y}^{\nu}) - (4amk)^2 \hat{y}^{\mu}\hat{y}^{\nu}\right]$ with $\hat{y}^{\alpha}$ a unit vector defined by $\hat{y}^{\alpha} = (0,0,1)$. It can be explicitly summed using the definition [7] $Li_n(x) = \sum_{k=1}^{\infty}\frac{x^k}{k^n}$ for the polylogarithm function $Li_n(x)$ and we get $\frac{1}{2\pi(4a)^3}\left[(g^{\mu\nu} - 3\hat{y}^{\mu}\hat{y}^{\nu})\Lambda + B\hat{y}^{\mu}\hat{y}^{\nu}\right]$, with

$\Lambda \equiv Li_3(e^{-4am}) + 4amLi_2(e^{-4am})$ and $B \equiv (4am)^2 \ln(1-e^{-4am})$. The second sum in (9) can also be reworked using a generalization of the polylogarithm function - namely the Lerch transcendent, it being defined by[7]

$\Phi(x,s,a) = \sum_{j=0}^{\infty}\frac{x^j}{(a+j)^s}$ giving for the sum $-\frac{1}{2\pi(4a)^3}\left\{\eta\left(\Psi(g^{\mu\nu} - 3\hat{y}^{\mu}\hat{y}^{\nu}) - 2\Delta\hat{y}^{\mu}\hat{y}^{\nu}\right) - 2g^{\mu\nu}(\Psi+\Delta)\right\}$,

with $\eta = \begin{cases} 1, & \nu=0,1 \\ -1, & \nu=2 \end{cases}$, $\Delta \equiv (4am)^2 \tanh^{-1}(e^{-2am})$ and $\Psi \equiv e^{-2am}\left(\Phi\left(e^{-4am}, 3, \frac{1}{2}\right) + 4am\Phi\left(e^{-4am}, 2, \frac{1}{2}\right)\right)$ .

Since $\Psi$ and $\Delta$ are non – negative; the Casimir energy density now becomes

$$\Theta^{00}_{vac}[\partial\Gamma] = \frac{1}{2\pi(4a)^3}(\Lambda + \Psi + 2\Delta) \quad (10)$$

which is non – negative.